\documentclass[%
 reprint,
 amsmath,amssymb,
 aps,
 pre,
 superscriptaddress
]{revtex4-1}

\usepackage{CJK}
\usepackage{graphicx}
\usepackage{dcolumn}
\usepackage{bm}
\usepackage{xcolor}
\usepackage{amsmath}
\usepackage{float}

\begin{document}

\begin{CJK*}{UTF8}{} 

\title{Stratification of drying particle suspensions: Comparison of implicit and explicit solvent simulations}

\author{Yanfei Tang ({\CJKfamily{gbsn}唐雁飞})}
\affiliation{Department of Physics, Center for Soft Matter and Biological Physics, and Macromolecules Innovation Institute, Virginia Polytechnic Institute and State University, Blacksburg, Virginia 24061, USA}
\author{Gary S. Grest}
\affiliation{Sandia National Laboratories, Albuquerque, NM 87185, USA}
\author{Shengfeng Cheng ({\CJKfamily{gbsn}程胜峰})}
\email{chengsf@vt.edu}
\affiliation{Department of Physics, Center for Soft Matter and Biological Physics, and Macromolecules Innovation Institute, Virginia Polytechnic Institute and State University, Blacksburg, Virginia 24061, USA}

\date{\today}

\begin{abstract}
Large scale molecular dynamics simulations are used to study drying suspensions of a binary mixture of large and small particles in explicit and implicit solvents. The solvent is first modeled explicitly and then mapped to a uniform viscous medium by matching the diffusion coefficients and the pair correlation functions of the particles. ``Small-on-top'' stratification of the particles, with an enrichment of the smaller ones at the receding liquid-vapor interface during drying, is observed in both models under the same drying conditions. With the implicit solvent model, we are able to model much thicker films and study the effect of the initial film thickness on the final distribution of particles in the dry film. Our results show that the degree of stratification is controlled by the P\'{e}clet number defined using the initial film thickness as the characteristic length scale. When the P\'{e}clet numbers of large and small particles are much larger than 1, the degree of ``small-on-top'' stratification is first enhanced and then weakens as the P\'{e}clet numbers are increased.
\end{abstract}

\maketitle

\end{CJK*}

\section{Introduction}

The stratifying phenomena in drying suspensions of polydisperse particles have recently attracted great interest since the clear demonstration of the counterintuitive ``small-on-top'' stratification by Fortini \textit{et al.},\cite{Fortini2016} where smaller particles are found to be enriched at the evaporating surface and distributed on top of larger particles after very fast drying. Since then, a burgeoning number of papers have appeared on the physical mechanisms underlying stratification\cite{Martin-Fabiani2016, Zhou2017, Fortini2017, Howard2017, Howard2017b, Makepeace2017, Sear2017, Sear2018, Liu2018, Tatsumi2018, Cusola2018, Tang2018Langmuir, Carr2018, Statt2018, Martin-Fabiani2018, Zhou2017b, Schulz2018} and possible approaches of its control.\cite{Martin-Fabiani2016, Tang2019Langmuir_control} The current physical picture for understanding stratification is based on the competition between the accumulation of particles at a receding liquid-vapor interface during evaporation and their diffusion away from the interface. This competition is quantified by a dimensionless P\'{e}clet number,\cite{Routh1998} $\text{Pe} = Hv_e/D$, where $H$ is a characteristic length scale and can be taken as the initial thickness of a drying film, $v_e$ is the speed at which the interface recedes, and $D$ is the diffusion coefficient of the particles. Particles with different diffusion coefficients thus have different P\'{e}clet numbers. One interesting polydisperse system is a suspension of a binary mixture of particles differing only in their diameters, $d_l$ for the larger ones and $d_s$ for the smaller ones. The size ratio is $\alpha = d_l /d_s > 1$. If the Stokes-Einstein relationship holds, then the ratio of the corresponding P\'{e}clet numbers is $\text{Pe}_l / \text{Pe}_s =\alpha$. The discovery of Fortini \textit{et al.} is that in the regime $\text{Pe}_l \gg \text{Pe}_s \gg 1$, ``small-on-top'' stratification occurs,\cite{Fortini2016} in contrast to ``large-on-top'' stratification for $\text{Pe}_l > 1 > \text{Pe}_s$ that was established earlier by Routh and collaborators.\cite{Routh2004, Trueman2012, Trueman2012a, Atmuri2012} The argument of Fortini \textit{et al.}\cite{Fortini2016} and Zhou \textit{et al.}\cite{Zhou2017} is that when evaporation is very fast, the numbers of large and small particles are initially both enhanced near the descending interface. This accumulation leads to a particle concentration gradient, which generates a phoretic driving force to push particles out of the particle-concentrated region. However, the driving force is asymmetric and the induced drifting velocity is larger for larger particles.\cite{Fortini2016, Zhou2017} As a result, a larger fraction of the bigger particles are pushed out of the region near the receding interface while relatively more of the smaller particles are left there, creating a ``small-on-top'' stratified state. This picture is termed the diffusiophoretic model of stratification.\cite{Sear2017, Sear2018}

The field of drying-induced stratification was recently reviewed by Schulz and Keddie.\cite{Schulz2018} Molecular modeling has played an important role in the process of discovering and revealing the fundamental physics of stratification.\cite{Fortini2016, Martin-Fabiani2016, Makepeace2017, Howard2017, Howard2017b, Tatsumi2018, Tang2018Langmuir, Statt2018} Fortini \textit{et al.} conducted Langevin dynamics simulations to unequivocally establish the occurrence of ``small-on-top'' stratification during fast drying in bidisperse particle suspensions.\cite{Fortini2016, Martin-Fabiani2016, Makepeace2017} Howard \textit{et al.} adopted a similar method and combined it with a dynamical density functional theory to show that ``small-on-top'' stratification is enhanced when the particle size ratio $\alpha$ is increased.\cite{Howard2017, Howard2017b} Tatsumi \textit{et al.} performed Langevin dynamics simulations for $\alpha =1.5$, 2, and 4 with particle-particle interactions described by the Hertzian theory of a nonadhesive elastic contact and showed that segregation is most enhanced at an intermediate value of $\text{Pe}_l$.\cite{Tatsumi2018} In all these studies, the solvent was treated as a continuous, viscous, and isothermal background with hydrodynamic flow ignored, which is consistent with the assumption usually made in phenomenological theories of stratification.\cite{Fortini2016, Zhou2017} However, the recent analyses of Sear and Warren showed that the solvent backflow around a migrating particle may be important and theories neglecting it may substantially overestimate stratification.\cite{Sear2017} The implication is that results based on implicit solvent models may not be realistic.\cite{Schulz2018}

Statt \textit{et al.} used molecular dynamics (MD) simulations to investigate stratification in drying mixtures of long and short polymer chains and compared the results from an implicit and an explicit solvent model.\cite{Statt2018} They carefully matched the sizes of polymer chains and their diffusion coefficients in the two models. With the implicit solvent, stratification was observed while no stratification occurred in the explicit solvent under the same drying conditions. They concluded that hydrodynamic interactions, which are not included in the implicit solvent model, are responsible for the different outcomes. The work by Statt \textit{et al.} thus presents a serious challenge to the modeling of drying particle suspensions as it raises a question whether one can trust the results from simulations based on implicit solvent models. It should be noted that these simulations are for polymer solutions and it is unclear if the results can be generalized to colloidal suspensions, though Statt \textit{et al.} suggested that they should apply to particle mixtures.\cite{Statt2018}

In our previous work,\cite{Tang2018Langmuir, Tang2019Langmuir_control} we have employed MD simulations with an explicit solvent modeled as a Lennard-Jones liquid to study the drying process of suspensions of bidisperse mixtures of nanoparticles. Though thermophoresis caused by evaporative cooling competed with diffusiophoresis and complicated the distribution of nanoparticles during drying, ``small-on-top'' stratification was observed, underscoring the discovery of Fortini \textit{et al.}. In this paper, we use a similar model but suppress thermophoresis by thermalizing the entire solvent and thus keeping the system isothermal during evaporation. Then we map the explicit solvent model to an implicit one by matching the diffusion coefficients of nanoparticles via tuning the frictional damping in the corresponding Langevin equation, as well as the pair correlation functions of nanoparticles by slightly adjusting their size parameters in the nanoparticle-nanoparticle interaction potentials in the implicit solvent. We compare the results from the explicit and implicit solvent models and find comparable ``small-on-top'' stratification in both. Our results thus corroborate the usage of an implicit solvent model for drying particle suspensions. Furthermore, we use the implicit solvent model to study the effect of initial thickness of a suspension film of nanoparticles on their final distribution in the dry film when either the P\'{e}clet number or the receding speed of the film's free surface is fixed.

\section{Model and Methodology}

We performed MD simulations with either an explicit or an implicit solvent model to study the drying process of suspensions containing a bidisperse mixture of nanoparticles. The explicit solvent model was described in detail in our previous study \cite{Tang2018Langmuir, Tang2019Langmuir_control} and is summarized below. The implicit solvent model is based on the method of Fortini \textit{et al.} \cite{Fortini2016} to mimic the process of solvent evaporation by moving the location of the liquid-vapor interface. We carefully matched the two models such that the particles have the same, purely repulsive interactions with each other, exhibit the same diffusive behavior, and have almost the same pair correlation functions in the explicit and implicit solvents. By comparing the results from these two models, we study the role of the solvent during drying. In particular, the possible effects of hydrodynamic interactions in drying particle suspensions, which are not captured by the implicit solvent model, will be clarified.

\subsection{Explicit solvent model} \label{ss:exp_method}

The explicit solvent is modeled as a fluid consisting of beads of mass $m$ that interact with each other via a LJ potential
\begin{equation} \label{eq:lj_potential}
U_{\text{LJ}}(r) = 4\epsilon \left[ \left(\frac{\sigma}{r}\right)^{12} - \left(\frac{\sigma}{r}\right)^6 - \left(\frac{\sigma}{r_c}\right)^{12} + \left(\frac{\sigma}{r_c}\right)^6 \right],
\end{equation}
where $r$ is the center-to-center separation between beads, $\epsilon$ is an energy unit, and $\sigma$ is a length unit. The potential is truncated at $r_c = 3 \sigma$ for the solvent. The nanoparticles are modeled as spheres with a uniform distribution of LJ mass points at density $1.0 m/\sigma^3$.\cite{Everaers2003,intVeld2008} The large nanoparticles (LNPs) have diameter $d_l = 20 \sigma$ and mass $m_l = 4188.8m$. The small nanoparticles (SNPs) have diameter $d_s = 5 \sigma$ and mass $m_s = 65.4m$. The size ratio is $\alpha = d_l/d_s = 4$. The nanoparticles interact with each other via an integrated LJ potential with a Hamaker constant $A_{\rm nn} = 39.48 \epsilon$.\cite{Everaers2003, intVeld2008} To avoid aggregation, the direct nanoparticle-nanoparticle interactions are made purely repulsive by truncating the potentials at their minima, which are $20.574 \sigma$ for LNP-LNP, $13.085\sigma$ for LNP-SNP, and $5.595\sigma$ for SNP-SNP pairs, respectively. The nanoparticle-solvent interaction is also modeled as an integrated LJ potential with a Hamaker constant $A_{\rm ns} = 100\epsilon$ and a cutoff $d/2 + 4\sigma$ with $d$ being the nanoparticle diameter.\cite{Everaers2003, intVeld2008} The nanoparticle-solvent interaction adopted here is strong enough to guarantee that both LNPs and SNPs are well dispersed in the solvent but not too strong to lead to solvent layers bound to the nanoparticles.\cite{Cheng2012}
  
\begin{table*}[ht]
\centering
\caption{Parameters for all systems studied.}
\begin{tabular}{ccccccccc}
\hline
System          & $H(0)/\sigma$ & $N_l$  & $N_s$  & $\phi_l$ & $\phi_s$ & $v_e\tau/\sigma$   & $\text{Pe}_l$ & $\text{Pe}_s$  \\ \hline
$H_e$           & 304       & 200    & 6400   & 0.068    & 0.034    & $1.18 \times 10^{-3}$ & 99.4        & 17.0     \\
$H_1 v_1$       & 304       & 200    & 6400   & 0.068    & 0.034    & $1.18 \times 10^{-3}$ & 99.4        & 17.0  \\ 
$H_2 v_{1/2}$   & 626.5     & 400    & 12800  & 0.066	  & 0.033    & $5.91 \times 10^{-4}$ & 102.5        & 17.5 \\
$H_4 v_{1/4}$  & 1246.5    & 800    & 25600  & 0.067    & 0.033    & $2.96 \times 10^{-4}$ & 102.2         & 17.5 \\
$H_8 v_{1/8}$ & 2476.5    & 1600   & 51200  & 0.067    & 0.033    & $1.50 \times 10^{-4}$ & 102.9         & 17.6   \\
$H_2 v_1$   & 626.5     & 400    & 12800  & 0.066	  & 0.033    & $1.18 \times 10^{-3}$ & 204.8        & 35.0 \\
$H_4 v_1$  & 1246.5    & 800    & 25600  & 0.067    & 0.033    & $1.18 \times 10^{-3}$ & 407.4         & 69.7 \\
$H_8 v_1$ & 2476.5    & 1600   & 51200  & 0.067    & 0.033    & $1.18 \times 10^{-3}$ & 809.5         & 138.5   \\
\hline
\end{tabular}
\label{tb:system}
\end{table*}  

A rectangular box with dimensions $L_x \times L_y \times L_z$ is used as the simulation cell, where $L_x = L_y= 201\sigma$ and $L_z$ is varied for each system. The liquid-vapor interface is in the $x$-$y$ plane, in which periodic boundary conditions are imposed. Two walls at $z = 0$ and $z = L_z$ are used to confine all particles in the cell. The particle-wall interaction is given by a LJ 9-3 potential,
\begin{eqnarray} \label{eq:wall_potential}
U_W (h) &=& \epsilon_W \left[ \frac{2}{15}\left(\frac{D_W}{h}\right)^9 - \left(\frac{D_W}{h}\right)^3 \right. \nonumber \\
&-& \left. \frac{2}{15}\left(\frac{D_W}{h_c}\right)^9 + \left(\frac{D_W}{h_c}\right)^3 \right]~,
\end{eqnarray}
where $\epsilon_W = 2.0\epsilon$ is the interaction strength, $D_W$ is a characteristic length, $h$ is the distance between the center of a particle and the wall, and $h_c$ is the cutoff of the potential. For the solvent, we set $D_W = 1\sigma$ and $h_c = 3\sigma$ ($0.8583\sigma$) at the lower (upper) wall. The lower wall is thus wetted by the solvent while the upper wall is nonwetted. For the nanoparticles, both walls are nonadsorptive with $D_W = d/2$ and $h_c = 0.8583D_W$, where $d$ is the nanoparticle diameter.

Prior to evaporation, the explicit solvent system has $L_z = 477\sigma$ and contains 200 LNPs, 6400 SNPs, and $7.1\times 10^6$ solvent beads. The system is well equilibrated with a liquid-vapor interface located at height $H(0)=304\sigma$. The volume fractions are $\phi_l = 0.068$ for LNPs and $\phi_s = 0.034$ for SNPs. The diffusion coefficients of nanoparticles in the equilibrium suspension are determined as $D_l = 3.61 \times 10^{-3} \sigma^2/\tau$ for LNPs and $D_s = 2.11 \times 10^{-2} \sigma^2/\tau$ for SNPs.\cite{Tang2019Langmuir_control} The ratio $D_s/D_l = 5.8$, which is larger than the size ratio $\alpha$. The deviation from the Stokes-Einstein relation is due to the finite concentrations of nanoparticles.\cite{intVeld2008, intVeld2009} To implement evaporation, a rectangular box with dimensions $L_x \times L_y \times 20\sigma$ from the top wall is designated as the deletion zone and $\zeta$ solvent beads are removed every $\tau$ from this zone. For this paper, $\zeta = 30$ to yield a very fast evaporation rate. At this rate, the liquid-vapor interface recedes at an almost constant speed $v_e = \left[ H(0)-H(t) \right]/t$, where $H(t)$ is the film thickness at time $t$ clocked since the initiation of evaporation. With $D_l$, $D_s$, $H(0)$, and $v_e$ known, the P\'{e}clet numbers for LNPs and SNPs, $\text{Pe}_l$ and $\text{Pe}_s$, are computed. All parameters are listed in Table~\ref{tb:system}.

All simulations were conducted with Large-scale Atomic/Molecular Massively Parallel Simulator (LAMMPS).\cite{Plimpton1995} The equation of motion is integrated by a velocity-Verlet algorithm with a time step $\delta t = 0.01 \tau$. A Langevin thermostat with a damping time $\Gamma = 100 \tau$ is used for the entire solvent including the vapor to fix its temperature at $1.0\epsilon/k_{\rm B}$. Therefore, the system is isothermal and thermophoresis is suppressed,\cite{Tang2018Langmuir, Tang2019Langmuir_control} as typically assumed in an implicit solvent model. Our previous study showed that this thermostat is weak enough that the screening effect on hydrodynamic interactions from the Langevin dynamics model adopted here is negligible.\cite{Tang2019Langmuir_control} Furthermore, almost the same stratification behavior was observed with the momentum-conserving dissipative particle dynamics thermostat.\cite{Tang2019Langmuir_control}

\subsection{Implicit solvent model} \label{ss:imp_method}

An implicit solvent system is prepared by removing all solvent from an equilibrated explicit solvent suspension. The strengths of nanoparticle-nanoparticle and nanoparticle-wall interactions remain unchanged. The role of the liquid solvent in the explicit model, in which the nanoparticles are suspended, is replaced by a potential barrier that confines all nanoparticles in the suspension. For each nanoparticle, the confining potential has the form of the right half of a harmonic potential and its minimum is always located at $d/2$ below the location of the liquid-vapor interface, where $d$ is the diameter of the nanoparticle. In other words, the contact angle of the nanoparticle is set as 0.\cite{Tang2018PRE} A nanoparticle experiences a Hookean restoring force that pushes it back into the suspension when the particle is near the interface. To mimic evaporation, the location of the liquid-vapor interface is moved downward along the $z$-axis, i.e., the instantaneous film thickness $H(t)$ is decreased at a given speed, $v_e$.\cite{Fortini2016, Martin-Fabiani2016, Makepeace2017, Howard2017, Howard2017b, Tatsumi2018, Statt2018} Therefore, $H(t) = H(0) - v_e t$ with $t$ being the time elapsed after the start of evaporation. Mathematically, the force exerted on the nanoparticle by the liquid-vapor interface is given by
\begin{eqnarray}
F^{i}_{z} = \left\{
	\begin{array}{ll}
-k_s \left[ z_n - H(t) + \frac{d}{2}\right]  & \text{for}~|z_n - H(t)| \le \frac{d}{2} \\
0   & \text{otherwise},
	\end{array}
\right.
\label{eq:confining}
\end{eqnarray}
where $k_s$ is a spring constant characterizing the strength of the confining potential, and $z_n$ is the nanoparticle position along the $z$ axis. Previously, we analyzed the capillary force experienced by a spherical particle adsorbed at a liquid-vapor interface,\cite{Tang2018PRE} which depends on the contact angle of the liquid on the particle surface. Our results show that the Hookean form in Eq.~(\ref{eq:confining}) is a reasonable approximation, though caution needs to be taken in the physical interpretation of $k_s$.\cite{Tang2018PRE} In this paper, we use $k_s = 0.3 \epsilon/\sigma^2$.

For all implicit solvent simulations, the times step $\delta t = 0.005 \tau$. A Langevin thermostat is applied to all nanoparticles in order to maintain the temperature of the system at $1.0\epsilon/k_{\rm B}$. To compare the two solvent models, we matched the diffusion coefficients of nanoparticles in the implicit solvent model to those in the explicit solvent. To this end, we tuned the damping time, $\Gamma$, of the Langevin thermostat applied to LNPs and SNPs in the implicit solvent. With $\Gamma =15.7\tau$ for LNPs and $1.53\tau$ for SNPs, the resulting diffusion coefficients of LNPs and SNPs in the implicit solvent are almost identical with those in the explicit solvent at the initial volume fractions.

\begin{figure}[ht]
\centering
\includegraphics[width=0.5\textwidth]{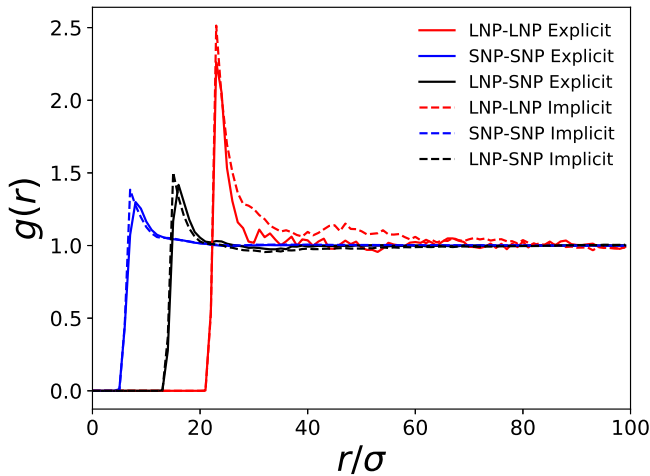}
\caption{Comparison of pair correlation functions of LNP-LNP (red, rightmost), SNP-SNP (blue, leftmost), and LNP-SNP (black, middle) pairs in the explicit (solid lines) and implicit (dashed lines) solvents.}
\label{fg:gofr}
\end{figure}

In the explicit solvent, there are solvent-mediated interactions between nanoparticles that effectively increase their sizes.\cite{Grest2011, Tang2018Langmuir, Tang2019Langmuir_control} To capture this effect, we slightly increased the diameter parameter of LNPs to $22.3\sigma$ and that of SNPs to $6.2\sigma$ in the nanoparticle-nanoparticle interaction potentials to ensure that their pair correlation functions in the explicit and implicit solvents are closely matched, as shown in Fig.~\ref{fg:gofr}.

Since an implicit solvent system only contains nanoparticles and is computationally much more efficient, we were able to study thicker suspension films and explore the effect on stratification of the initial film thickness, $H(0)$, with the initial volume fractions of LNPs and SNPs fixed. The value of the receding speed of the liquid-vapor interface, $v_e$, is either fixed, where the P\'{e}clet numbers increase proportionally with the initial thickness of a film, or varied to yield similar P\'{e}clet numbers as in the system with $H(0) =304\sigma$ and $v_e = 1.18 \times 10^{-3}\sigma/\tau$.

All systems studied in this paper are summarized in Table~\ref{tb:system}. The number of LNPs is $N_l$ and that of SNPs is $N_s$. We use $H_e$ to denote the explicit solvent system, which has $H(0) =304\sigma$ and $v_e = 1.18 \times 10^{-3}\sigma/\tau$. The implicit solvent system with the same initial film thickness and evaporation rate is denoted as $H_1 v_1$. For other implicit solvent systems, $H_q v_f$ is used to indicate that the initial film thickness is $q\times H(0)$ and the receding speed of the interface is $v_e = f\times 1.18 \times 10^{-3}\sigma/\tau$. In this paper, we vary $q$ from 1 to 8 and $f$ from 1 to 1/8.

\begin{figure}[h]
\centering
\includegraphics[width=0.5\textwidth]{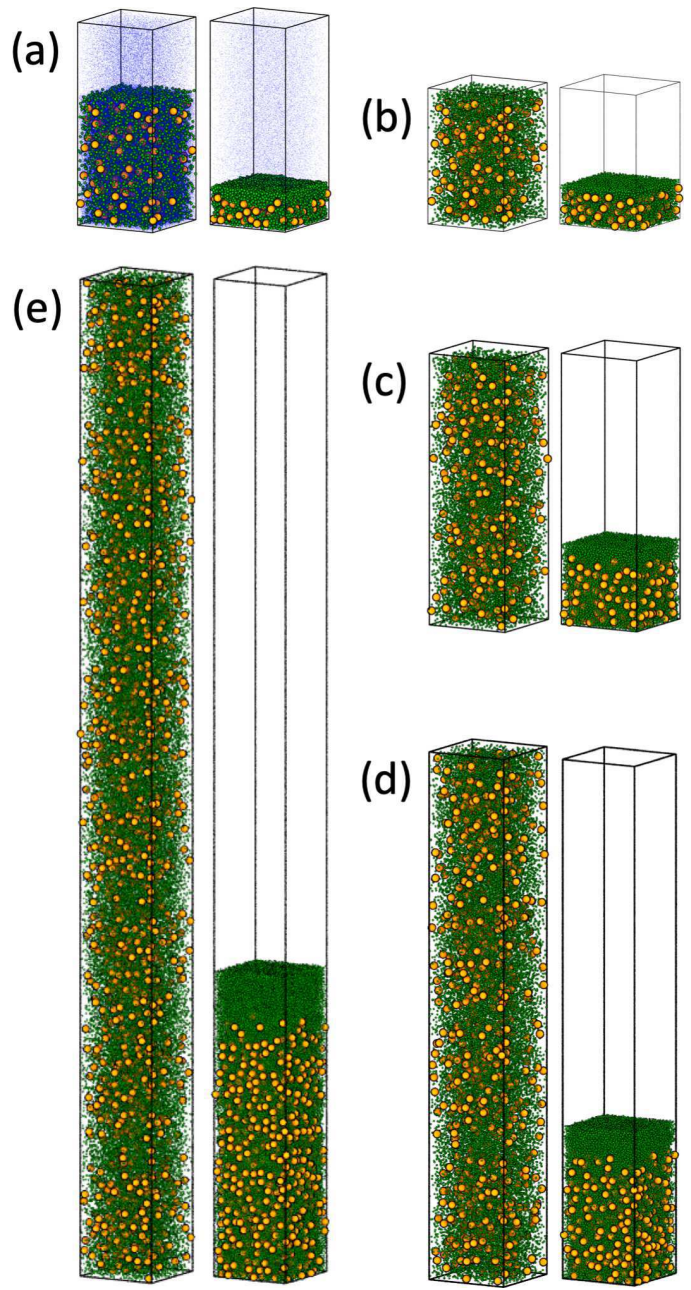}
\caption{Snapshots of drying suspensions: (a) the explicit solvent system $H_e$; the implicit solvent systems (b) $H_1 v_1$, (c) $H_2v_{1/2}$, (d) $H_4v_{1/4}$, and (e) $H_8v_{1/8}$. For each system, the left snapshot is for the equilibrium suspension prior to evaporation while the right one is for the state with $H(t)\simeq 0.3 H(0)$. Color code: LNPs (orange), SNPs (green), and solvent (blue). For $H_e$, only 5\% of the solvent beads are visualized to improve clarity.}
\label{fg:snaps}
\end{figure}

\section{Results and Discussion}

\begin{figure*}[htb]
\centering
\includegraphics[width=\textwidth]{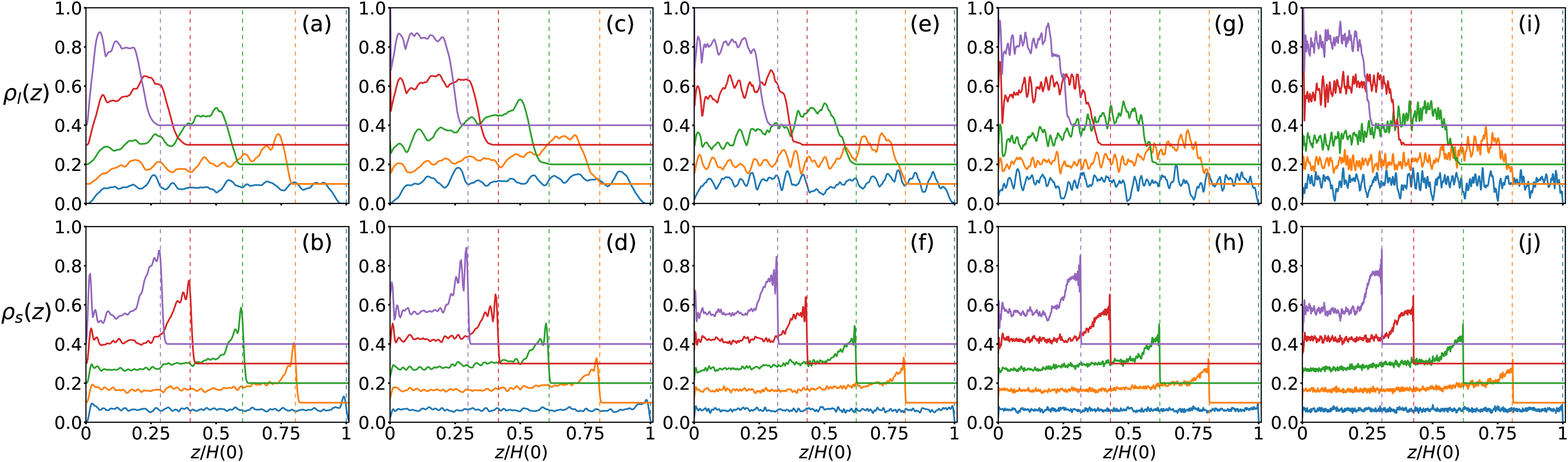}
\caption{Evolution of density profiles for LNPs (top row) and SNPs (bottom row) for $H_e$ [(a) and (b)], $H_1v_1$ [(c) and (d)], $H_2v_{1/2}$ [(e) and (f)], $H_4v_{1/4}$ [(g) and (h)], and $H_8v_{1/8}$ [(i) and (j)]. The vertical dashed lines indicate the location of the liquid-vapor interface. For clarity, each profile is shifted vertically by $0.1m/\sigma^3$ from the previous one.}
\label{fg:density}
\end{figure*}

In Fig.~\ref{fg:snaps} we show snapshots of five suspensions before and after drying, including $H_e$, $H_1 v_1$, $H_2v_{1/2}$, $H_4v_{1/4}$, and $H_8v_{1/8}$ that all have similar P\'{e}clet numbers. After the film thickness is reduced to $H(t)\simeq 0.3 H(0)$, all systems exhibit ``small-on-top'' stratification. The ``small-on-top'' stratified state is more visually prominent for thicker films such as $H_2v_{1/2}$, $H_4v_{1/4}$, and $H_8v_{1/8}$, though quantitative analyses show that the degree of stratification in these thick films is close to that in $H_e$. Furthermore, $H_e$ and $H_1v_1$ have identical distributions of nanoparticles prior to evaporation and are dried at very similar rates. At $H(t) \simeq 0.3 H(0)$ when the simulations were stopped, $H_e$ exhibits slightly stronger ``small-on-top'' stratification than $H_1v_1$, which may be due to the fact in the explicit solvent, the diffusion coefficients of nanoparticles decrease as their concentration increases during solvent evaporation. This observation is in discordance with the theoretical analysis of Sear and Warren\cite{Sear2017} and the simulation study of Statt \textit{et al.}\cite{Statt2018} Sear and Warren showed that the back-flow around a drifting particle in an explicit solvent suppresses the diffusiophoretic driving on the larger particles from a concentration gradient of the smaller particles.\cite{Sear2017} As a result, ``small-on-top'' stratification is expect to be significantly promoted in an implicit solvent model where back-flow is missing.\cite{Sear2017} Statt \textit{et al.} used MD to simulate a mixture of long and short polymer chains in an explicit and an implicit solvent and found that the implicit solvent system exhibits ``small-on-top'' stratification, whereas the explicit one does not.\cite{Statt2018} However, the analysis of Sear and Warren\cite{Sear2017} is based on the Asakura-Oosawa model,\cite{Asakura1954} which is about the diffusion of a very large particle in a polymer solution with concentration gradients. The simulations of Statt \textit{et al.} are for polymer mixtures.\cite{Statt2018} We suspect that colloidal suspensions and polymer solutions behave quite differently in terms of diffusiophoresis and stratification. It is interesting to explore if the Asakura-Oosawa model can be extended to a particle with size comparable to the sizes of polymer chains in the solution, where the curvature of the particle comes into play.

For quantitative analyses, in Fig.~\ref{fg:density} we plot the density profiles of LNPs and SNPs along the normal direction of the film, which are computed as $\rho_{i}(z) = n_{i} (z) m_i / (\sigma L_x L_y)$ with $i \in \{l,s\}$. Specifically, $n_{i} (z)$ is the number of $i$-type particles in a spatial bin of thickness $\sigma$ centered on $z$ and $m_i$ is the mass of one $i$-type particle. For a nanoparticle straddling several bins, its contribution to $n_{i} (z)$ is a fraction equal to the ratio between its volume enclosed by each bin and the entire volume of the nanoparticle. To compare different films, in Fig.~\ref{fg:density} we normalize $z$ by the initial film thickness, $H(0)$, for each suspension film.

Several features can be easily identified from these density profiles. During drying, both LNPs and SNPs are enriched near the receding liquid-vapor interface since ${\rm Pe}_l \gg {\rm Pe}_s \gg 1$ and all five systems exhibit qualitatively similar density profiles. However, the enrichment of SNPs in the interfacial region is stronger in its degree than that of LNPs. For all implicit solvent systems, the density profiles at the same stage of drying (i.e., at the same $H(t)/H(0)$) are all similar. In the final state with $H(t)\simeq 0.3 H(0)$, the density profile of LNPs along the $z$-axis has a slight negative gradient for $H_e$, is almost flat for $H_1v_1$ and $H_2v_{1/2}$, while exhibits a very weak positive gradient for $H_4v_{1/4}$ and $H_8v_{1/8}$. Therefore, $H_e$ with an explicit solvent is expected to display stronger ``small-on-top'' stratification than all implicit solvent systems, while stratification of similar amplitudes is expected for $H_4v_{1/4}$ and $H_8v_{1/8}$.

The state of stratification can be characterized by examining the mean heights of LNPs and SNPs as a function of time, which are computed as $\langle z_i \rangle= \frac{1}{N_i} \sum\limits_{n=1}^{N_i} z_{i,n}$ with $i \in \{l, s\}$ for LNPs and SNPs, respectively. Here $z_{i,n}$ is the $z$ coordinate of the $n$-th nanoparticle of type $i$. An order parameter of stratification can then be defined as $ (2\langle z_l \rangle -2 \langle z_s\rangle)/H(t) $, i.e., as the difference in the average height of LNPs and that of SNPs normalized by a half of the instantaneous thickness of the drying film.\cite{Tang2018Langmuir} In the equilibrium suspension before evaporation, $ \langle z_l \rangle \simeq \langle z_s \rangle \simeq H(0)/2$ and therefore $ \langle z_l \rangle - \langle z_s\rangle \simeq 0$. After drying is initiated, $ \langle z_l \rangle - \langle z_s\rangle > 0$ indicates ``large-on-top'' stratification while  ``small-on-top'' corresponds to $ \langle z_l \rangle - \langle z_s\rangle < 0$.

\begin{figure}[h]
\centering
\includegraphics[width = 0.4\textwidth]{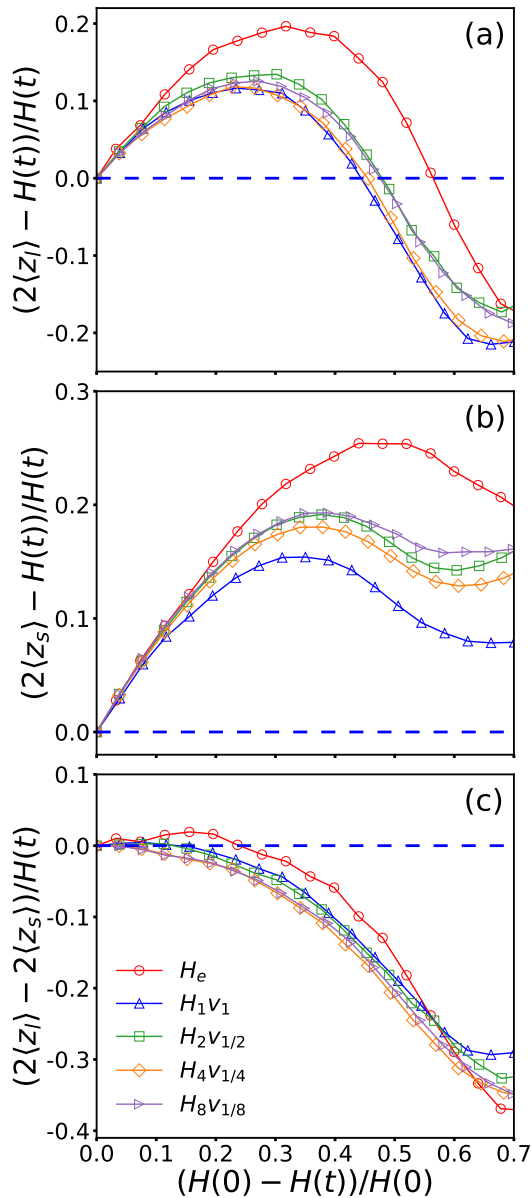}
\caption{Mean height relative to the center of a drying film of (a) LNPs and (b) SNPs, and (c) the difference in the average height of LNPs and SNPs, all normalized by $H(t)/2$, are plotted against the extent of drying, $(H(0)-H(t))/H(0)$, for $H_e$ (red circle), $H_1 v_1$ (blue upward triangle), $H_2v_{1/2}$ (green square), $H_4v_{1/4}$ (yellow diamond), and $H_8v_{1/8}$ (purple right-pointing triangle).}
\label{fg:op}
\end{figure}

In Fig.~\ref{fg:op} we plot $ (2\langle z_l \rangle -H(t))/H(t) $, $ (2\langle z_s \rangle -H(t))/H(t) $, and $ (2\langle z_l \rangle -2 \langle z_s\rangle)/H(t) $ against $1-H(t)/H(0)$ that quantifies the extent of drying. As shown in Fig.~\ref{fg:op}(a), the data on $\langle z_l \rangle$ are close together for all implicit solvent systems $H_1 v_1$, $H_2v_{1/2}$, $H_4v_{1/4}$, and $H_8v_{1/8}$ with similar P\'{e}clet numbers. In the early stage of the drying process, $\langle z_l \rangle$ is larger than $H(t)/2$, indicating that the LNPs are enriched in the top half of the drying film. However, in the late stage of drying, $\langle z_l \rangle -H(t)/2 < 0$ and the LNPs become more concentrated in the bottom half of the drying film. For the implicit solvent systems, this transition occurs at $H(t)/H(0) \simeq 0.55$. For $H_e$, a similar transition occurs slightly later at $H(t)/H(0) \simeq 0.45$. However, in the very late stage of drying at $H(t)/H(0) \simeq 0.3$, the relative height of LNPs with respect to the drying film is similar for all explicit and implicit solvent systems.

As shown in Fig.~\ref{fg:op}(b), the SNPs are always accumulated in the top half of the drying film for both explicit and implicit solvent models. When $H(t)/H(0) \lesssim 0.9$, $\langle z_s \rangle$ shows larger variations among the implicit solvent systems with different $H(0)$. The accumulation of SNPs in the top half of the drying film is weaker in $H_1 v_1$ and is enhanced when the initial film gets thicker. For $H_2v_{1/2}$, $H_4v_{1/4}$, and $H_8v_{1/8}$, the results of $ (2\langle z_s \rangle -H(t))/H(t) $ against $1-H(t)/H(0)$ are close to each other and the shift from one curve to another is nonmonotonic when $H(0)$ is increased (i.e., for $H_2v_{1/2} \rightarrow H_4v_{1/4} \rightarrow H_8v_{1/8}$), indicating that the initial films are thick enough to lead to a convergence in the behavior of SNPs. From Fig.~\ref{fg:op}(b), we also note that the accumulation of SNPs in the top half of the drying film is always stronger in the explicit solvent than in the implicit solvent.

From $ \langle z_l \rangle $ and $\langle z_s\rangle$, we expect that in the final dry film, $H_e$ should yield the strongest ``small-on-top'' stratification while $H_1 v_1$ should lead to the weakest. Furthermore, $H_2v_{1/2}$, $H_4v_{1/4}$, and $H_8v_{1/8}$ are expected to be very similar in terms of the degree of stratification. The plots of the order parameter of stratification in Fig.~\ref{fg:op}(c), $ (2\langle z_l \rangle -2 \langle z_s\rangle)/H(t) $ against $1-H(t)/H(0)$, confirm all these predictions. Our results clearly demonstrate the emergence of ``small-on-top'' stratification with comparable amplitudes in both explicit and implicit solvent models with similar P\'{e}clet numbers. Furthermore, the data confirm that the initial film thickness is the appropriate length scale entering the P\'{e}clet number.

\begin{figure*}[htb]
\centering
\includegraphics[width=\textwidth]{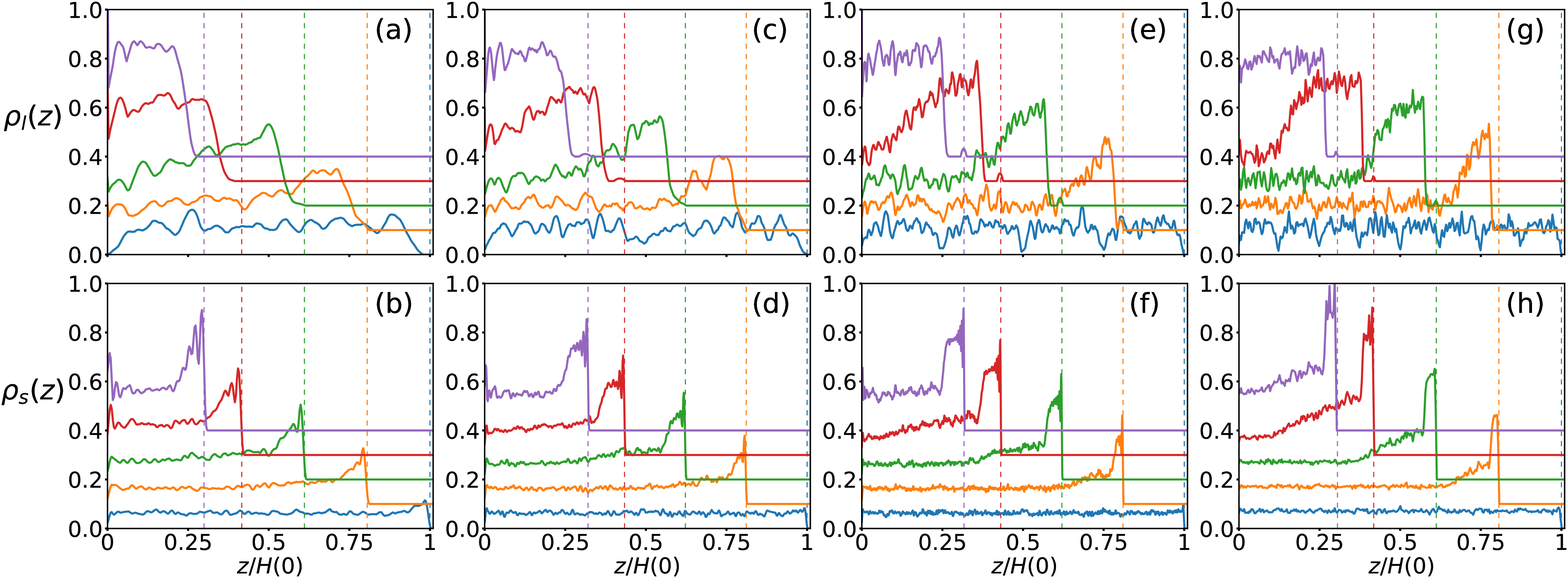}
\caption{Evolution of density profiles for LNPs (top row) and SNPs (bottom row) for $H_1v_1$ [(a) and (b)], $H_2v_1$ [(c) and (d)], $H_4v_1$ [(e) and (f)], and $H_8v_1$ [(g) and (h)]. The vertical dashed lines indicate the location of the liquid-vapor interface. For clarity, in each plot a profile is shifted upward by $0.1m/\sigma^3$ from the previous one.}
\label{fg:density_v1}
\end{figure*}

In contrast to the previous report of Statt \textit{et al.} on polymer solutions where ``small-on-top'' stratification only occurs in the implicit solvent system,\cite{Statt2018}  ``small-on-top'' occurs in both models here, with the degree of stratification comparable or even slightly stronger in the explicit solvent. Our results indicate that the physics of drying may have some differences in colloidal suspensions and polymer solutions. To map a polymer solution in an explicit solvent to a system with an implicit solvent, both the monomer-monomer interactions and the viscous damping on the monomers have to be adjusted to match the size (i.e., the radius of gyration) and diffusion of polymer chains. For a colloidal suspension in which the particles are well dispersed, we just need to tune the damping drag to match their diffusion coefficients and slightly adjust the size parameter in the integrated LJ potentials describing the particle-particle interactions to match their pair correlation functions. Statt \textit{et al.} concluded that hydrodynamic interactions are not captured by the implicit solvent model and their missing leads to the occurrence of ``small-on-top'' stratification in their polymer solutions with the implicit solvent.\cite{Statt2018} In nanoparticle suspensions studied here, hydrodynamic interactions seem to play a much weaker role but more work is needed to elucidate their possible effects.

Using the implicit solvent model, we have also studied the effect of increasing the initial film thickness at a fixed evaporation rate. We compare four systems, $H_1v_1$, $H_2v_1$, $H_4v_1$, and $H_8v_1$, where $H(0)$ is increased from $304\sigma$ to $2476.5\sigma$ but $v_e$ is fixed at $1.18\times 10^{-3}\sigma/\tau$. As results, the P\'{e}clet numbers increase proportionally with $H(0)$ and $\text{Pe}_l$ increases from 99.4 in $H_1v_1$ to 809.5 in $H_8v_1$. As shown in the bottom row of Fig.~\ref{fg:density_v1}, the density profiles of SNPs plotted against $z/H(0)$ are qualitatively similar for the four systems. The main difference is that the peak value of $\rho_s(z)$ at the evaporating interface becomes slightly larger for larger $H(0)$. Another difference is the appearance of a plateau of $\rho_s(z)$ just right below the highly SNP-enriched skin layer at the evaporating liquid-vapor interface when the film is thick enough, as in $H_2v_1$, $H_4v_1$, and $H_8v_1$. The absolute thickness of this plateau zone increases as $H(0)$ is increased, possibly indicating a jammed state of SNPs in this zone.\cite{Sear2018} Below this plateau, $\rho_s(z)$ first decreases sharply in a very narrow region and then gradually decreases as $z$ gets smaller, i.e., when it is further away from the evaporating front. Eventually, $\rho_s(z)$ reaches another plateau corresponding to the density of SNPs in the equilibrium suspension prior to evaporation. Fig.~\ref{fg:density_v1} also shows that the density profiles of LNPs remain qualitatively unchanged when $H(0)$ is increased (especially when $H(0)$ is large as in $H_4v_1$ and $H_8v_1$) but the evaporation rate is fixed. Going from the evaporating interface to the bulk of the drying suspension, $\rho_l(z)$ first decreases gradually and then decays rapidly to its value in the equilibrium suspension before evaporation. As $H(0)$ is increased, the peak value of $\rho_l(z)$ also becomes slightly larger (see the top row of Fig.~\ref{fg:density_v1}).

\begin{figure}[tb]
\centering
\includegraphics[width = 0.4\textwidth]{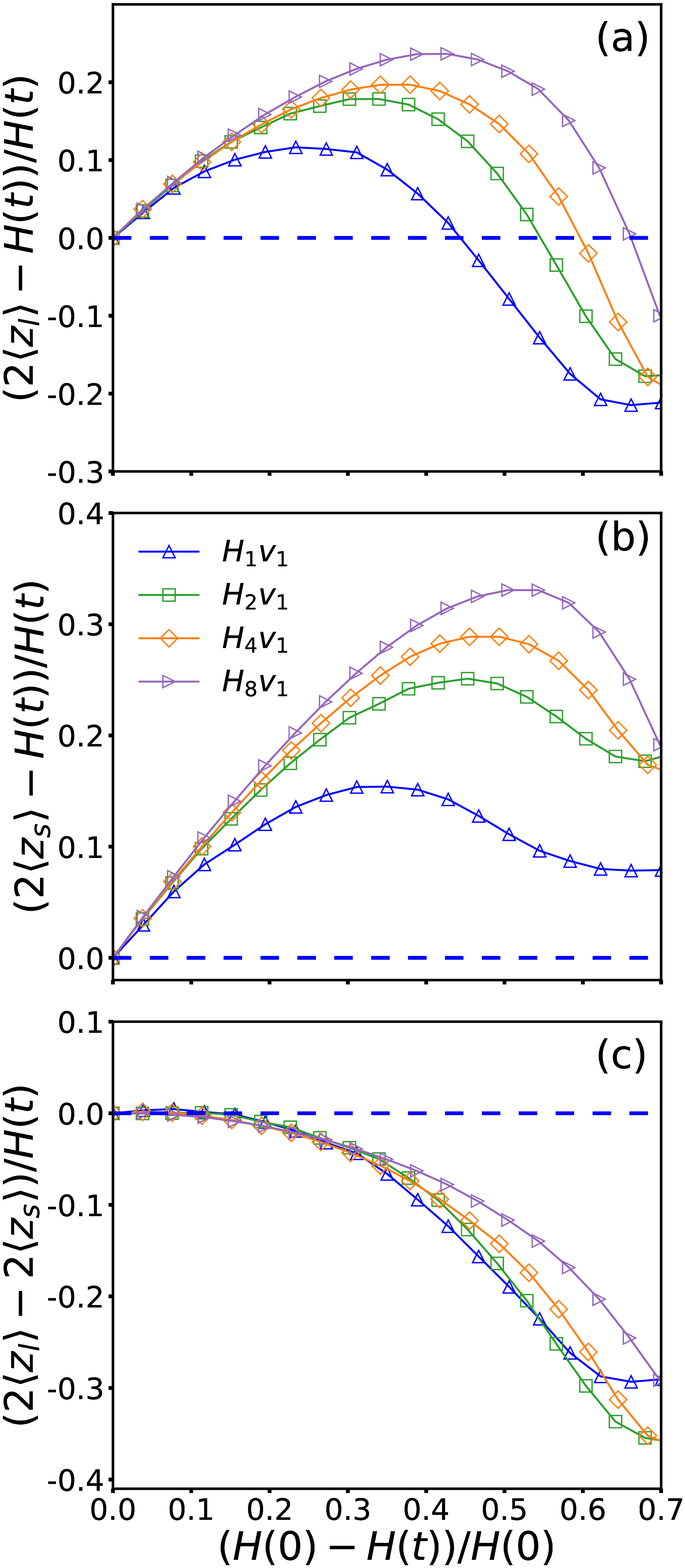}
\caption{Mean height relative to the center of a drying film of (a) LNPs and (b) SNPs and (c) the difference in the average height of LNPs and SNPs, all normalized by $H(t)/2$, are plotted against the extent of drying, $(H(0)-H(t))/H(0)$, for $H_1 v_1$ (blue upward triangle), $H_2v_1$ (green square), $H_4v_1$ (yellow diamond), and $H_8v_1$ (purple right-pointing triangle).}
\label{fg:op_v1}
\end{figure}

The average height of LNPs and SNPs plotted in Fig.~\ref{fg:op_v1}(a) and (b) shows interesting systematic changes as $H(0)$ is increased. First, $\langle z_s\rangle / H(t)$ shifts upward more considerably than $\langle z_l\rangle / H(t)$ with increasing $H(0)$. The data indicate that for a film with a larger initial thickness, the accumulation of both SNPs and LNPs near the receding interface and in the top half of the drying film is enhanced in the early stage of drying. As evaporation proceeds, the LNPs become more concentrated in the bottom half of the drying film and are at deficit in the top half, signaling ``small-on-top'' stratification. The transition between the enrichment of LNPs in the top half to their pileup in the bottom half occurs at a later stage of drying when $H(0)$ is increased, as shown in Fig.~\ref{fg:op_v1}(a). For all systems, Fig.~\ref{fg:op_v1}(b) shows that the SNPs are always accumulated in the top half of the drying film in the entire range of drying. As $H(0)$ increases, the SNPs form a thicker jammed layer below the receding interface and both SNPs and LNPs are trapped in this layer,\cite{Sear2018} though they are expected to be pushed out of the region close to the interface via the diffusiophoretic mechanism. The jamming effect may be underlying the observation that early on during drying there is enhanced enrichment of both SNPs and LNPs near the evaporating interface as $H(0)$ is increased and the accumulation of LNPs in the bottom half of the drying film arises later when $H(0)$ is larger. The order parameter of stratification in Fig.~\ref{fg:op_v1}(c) confirms that all four systems display ``small-on-top'' stratification, which emerges almost instantaneously once the solvent evaporation is initiated.

\begin{figure}[htb]
\centering
\includegraphics[width = 0.4\textwidth]{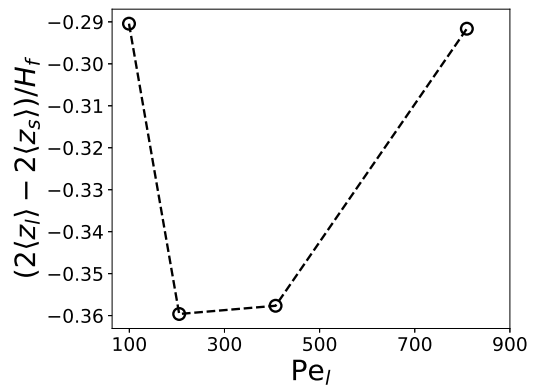}
\caption{Difference in the average height of LNPs and SNPs, normalized by $H_f/2$, is plotted against $\text{Pe}_l$ for $H_1 v_1$, $H_2v_1$, $H_4v_1$, and $H_8v_1$ from left to right. The data are extracted from Fig.~\ref{fg:op_v1}(c) at $H(t) =H_f \simeq 0.3H(0)$.}
\label{fg:op_peclet}
\end{figure}

When $v_e$ is fixed, the P\'{e}clet numbers become larger as $H(0)$ is increased. In Fig.~\ref{fg:op_peclet} we plot the amplitude of stratification $ (2\langle z_l \rangle -2 \langle z_s\rangle)/H(t) $ at $H(t) =H_f \simeq 0.3H(0)$ as a function of $\text{Pe}_l$. The four data points are for $H_1v_1$, $H_2v_1$, $H_4v_1$, and $H_8v_1$, respectively, at the same extension of drying. Note that a more negative value of $ (2\langle z_l \rangle -2 \langle z_s\rangle)/H(t) $ indicates stronger ``small-on-top'' stratification. Fig.~\ref{fg:op_peclet} shows that stratification is most pronounced for an intermediate value of $\text{Pe}_l$, which is around 300 for the systems studied here. This nonmonotonic behavior of the degree of stratification was also found in the simulations of Tatsumi \textit{et al.},\cite{Tatsumi2018} where the P\'{e}clet numbers were increased by increasing $v_e$ while fixing $H(0)$.

\section{Conclusions}

In this paper, we employ MD simulations to compare an explicit solvent model to an implicit one in studying the drying process of bidisperse particle suspensions. In the explicit model, the solvent is modeled as a Lennard-Jones liquid. In the implicit model, the solvent is treated as a viscous, uniform, isothermal background. In contrast to a previous report on polymer solutions where ``small-on-top'' stratification does not occur in the explicit solvent but occurs in the implicit one,\cite{Statt2018} we have observed the occurrence of comparable ``small-on-top'' stratification in both models. Our results indicate that the implicit solvent model can be used effectively for modeling the drying of thin film suspensions, for which the evaporative flow of the solvent is essentially one-dimensional. However, it remains unclear why the back-flow of the solvent around a migrating particle and the hydrodynamic interactions between the particles seem to be unimportant in the systems studied here.\cite{Sear2017, Statt2018}

With the implicit solvent model, we further study the effect of the initial film thickness on the drying of a suspension film of a bidisperse mixture of nanoparticles. Our results indicate that for films that are initially thick enough, the P\'{e}clet number is a valid dimensionless number capturing the competition between solvent evaporation and nanoparticle diffusion. For fast drying, the accumulation of either large or small nanoparticles near the receding interface is similar when the receding speed of the liquid-vapor interface is decreased in proportion to the increase of the initial film thickness, which results in similar P\'{e}clet numbers. For these systems, the degree of stratification is also similar. However, if the receding speed of the interface is fixed, then the accumulation near the interface is more significant for both large and small nanoparticles when the film gets thicker. The degree of stratification varies nonmonotonically and is most enhanced at an intermediate value of of the P\'{e}clet number, with $\text{Pe}_l \sim 300$ for the systems reported here.

In the systems studied here, the direct nanoparticle-nanoparticle interactions are purely repulsive in both solvent models to ensure that the nanoparticles are well dispersed in the suspension, though there might be solvent-mediated weak attractions between the nanoparticles in the explicit solvent. If there are direct attractions between the nanoparticles or strong nanoparticle-solvent attractions leading to a layer of solvent bound to each particle, then mapping an explicit solvent system to an implicit one requires a careful tuning of the nanoparticle-nanoparticle potentials to mimic the effect of the solvent. This mapping can be achieved by following the procedure outlined by Grest \textit{et al.} to derive an effective potential between nanoparticles in an implicit solvent.\cite{Grest2011} For such systems, it is an interesting question whether similar stratification can be observed in drying polydisperse particle suspensions with explicit and implicit solvents.

\section*{Acknowledgments}
Acknowledgment is made to the Donors of the American Chemical Society Petroleum Research Fund (PRF \#56103-DNI6), for support of this research. This research used resources of the National Energy Research Scientific Computing Center (NERSC), a U.S. Department of Energy Office of Science User Facility operated under Contract No. DE-AC02-05CH11231. These resources were obtained through the Advanced Scientific Computing Research (ASCR) Leadership Computing Challenge (ALCC). This work was performed, in part, at the Center for Integrated Nanotechnologies, an Office of Science User Facility operated for the U.S. Department of Energy Office of Science. Sandia National Laboratories is a multimission laboratory managed and operated by National Technology and Engineering Solutions of Sandia, LLC., a wholly owned subsidiary of Honeywell International, Inc., for the U.S. Department of Energy's National Nuclear Security Administration under contract DE-NA0003525. This paper describes objective technical results and analysis. Any subjective views or opinions that might be expressed in the paper do not necessarily represent the views of the U.S. Department of Energy or the United States Government.


\end{document}